\documentclass[proceedings]{JHEP37}
\usepackage{eurosym}
\usepackage{amssymb}
\usepackage{amsfonts}
\usepackage{amsmath,epsfig}
\usepackage{graphicx}

\newbox\mybox

\newcommand\fverb{\setbox\mybox=\hbox\bgroup\verb}
\newcommand\fverbdo{\egroup\medskip\noindent\fbox{\unhbox\mybox}\ }
\newcommand\fverbit{\egroup\item[\fbox{\unhbox\mybox}]}
\conference{Time-delay and reality conditions for complex solitons}
\abstract{We compute lateral displacements and time-delays for a scattering processes of complex multi-soliton solutions of the Korteweg de-Vries equation.
The resulting expressions are employed to explain the precise distinction between solutions obtained from different techniques, Hirota's direct method and a superposition principle based on B\"{a}cklund transformations. 
Moreover they explain the internal structures of degenerate compound multi-solitons previously constructed. Their individual one-soliton constituents are 
time-delayed when scattered amongst each other. We present generic formulae for these time-dependent displacements. By recalling Gardner's transformation method for conserved charges, we argue that the structure of the asymptotic behaviour resulting from the integrability of the model together
with its $\mathcal{PT}$-symmetry ensure the reality of all of these charges, including in particular the mass, the momentum and the energy.}

\title{Time-delay and reality conditions for complex solitons}
\author{Julia Cen$^\bullet$ and Francisco Correa$^{\dag,\ddag}$ and Andreas
Fring$^\bullet$ \\
$\bullet$ Department of Mathematics, City University London,\\
$\,\,$ Northampton Square, London EC1V 0HB, UK\\
$\dag$ Instituto de Ciencias F{\'{\i}}sicas y Matem{\'{a}}ticas, Universidad
Austral de Chile, \\
$\,\,$ Casilla 567, Valdivia, Chile\\
$\ddag$ Institut f{\"{u}}r Theoretische Physik and Riemann Center for
Geometry and Physics, \\
$\,\,$ Leibniz Universit{\"{a}}t Hannover, Appelstra\ss e 2, 30167 Hannover,
Germany\\
E-mail:
julia.cen.1@city.ac.uk,francisco.correa@itp.uni-hannover.de,a.fring@city.ac.uk%
}

\input{tcilatex}
\begin{document}

\section{Introduction}

It is one of the defining features of classical multi-soliton solutions to
nonlinear integrable equations that individual one-soliton contributions
maintain their overall shape before and after a scattering event. The only
net effect is that they are delayed or advanced in time as a result of the
scattering with other solitons when compared to the undisturbed motion of a
single one-soliton \cite{rubinstein,jackiw,fring1994vertex}. Besides
providing a more detailed picture on the motion of classical solitons,
following ideas of Wigner and Eisenbud \cite{wignereisen}, the concrete
values of the delay times are also important for the quantization of the
theory as they can be related to quantum mechanical scattering matrices in a
semi-classical approximation.

We present here a detailed analysis of the delay times for complex soliton
solutions to the Korteweg de-Vries equation (KdV) previously reported in 
\cite{CenFring,FranciscoAndreas}. We use our results for a variety of
purposes. On a technical level the explicit expressions allow to clarify the
precise distinctions between solutions obtained from different types of
solution methods, in particular Hirota's direct method and B\"{a}cklund
transformations. Moreover, the time-delays also shed new on the degenerate
multi-soliton solutions constructed in \cite{FranciscoAndreas}, especially
the internal structure of compound multi-solitons can be explained in detail
when using the expression for the time-delays or some approximate asymptotic
formulae.

In the quantum mechanical context it is well understood which role $\mathcal{%
PT}$-symmetries, or better antilinear symmetries \cite{EW}, play in order to
explain the reality of the energy eigenspectrum \cite%
{Bender:1998ke,Benderrev,Alirev}. For nonlinear integrable wave equations
and their $\mathcal{PT}$-symmetric deformations \cite%
{BBCF,AFKdV,Josh,BBAF,PEGAAFint,comp1,CompPaulo,CFB,YanZ,ACAF2,assis2015pt}
there is still some uncertainty about the precise reasoning. For the model
discussed here we argue that physical quantities based on one-soliton
solutions, $\mathcal{PT}$-symmetric or not, can always be made $\mathcal{PT}$%
-symmetric and in this case that feature alone guarantees their reality. The
integrability of the model then ensures that asymptotically any
multi-soliton solution separates into a collection of one-soliton solutions,
possibly time-delayed, of which each contributes only a real value to an
overall conserved charge. We recall the structure of all of these charges
constructed from Gardner's transformation and then use it to show that $%
\mathcal{PT}$-symmetry and integrability ensure the reality of all conserved
charges.

Our manuscript is organized as follows: Starting from some general
definitions and properties of conserved quantities and complex one-soliton
solutions we compute in section 2 the time-delays in nondegenerate and
degenerate two and three-soliton solutions. We provide closed expressions
for the time-dependent displacements for degenerate $N$-solitons for any $N$%
. In section 3 we present the reasoning that ensures the reality of all
conserved quantities. Our conclusions and an outlook into future work and
open issues is presented in section 4. In general the detailed computations
are omitted in section 2, but in order to illustrate the working we present
some sample computations in an appendix.

\section{Time-delays for multi-soliton solutions}

\subsection{Generalities}

Following \cite{rubinstein,jackiw,fring1994vertex} the classical time-delay
of a scattering process is defined as follows: We consider the trajectories
of a particle, or a soliton for that matter, with velocity $v$ before and
after the collision as $x=vt+x^{(i)}$ and $x=vt+x^{(f)}$, respectively. The 
\textit{lateral displacement} resulting from the scattering event is then
defined as the difference between these two trajectories, that is simply 
\begin{equation}
\Delta _{x}:=x^{(f)}-x^{(i)},  \label{DX}
\end{equation}%
so that the \textit{time-delay} is naturally defined as%
\begin{equation}
\Delta _{t}:=t^{(f)}-t^{(i)}=-\frac{x^{(f)}}{v}+\frac{x^{(i)}}{v}=-\frac{%
\Delta _{x}}{v}.  \label{DT}
\end{equation}%
Negative and positive time-delays are interpreted as attractive and
repulsive forces, respectively. In a multi-particle scattering process of
particles, or solitons, of type $k$ the corresponding lateral displacements
and time-delays $(\Delta _{x})_{k}$ and $(\Delta _{t})_{k}$, respectively,
have to satisfy certain consistence conditions \cite{fring1994vertex}.
Demanding for instance that the centre of mass coordinate%
\begin{equation}
X=\frac{\dsum\nolimits_{k}m_{k}x_{k}}{\dsum\nolimits_{k}m_{k}}
\end{equation}%
remains the same before and after the collision, i.e. $X^{(i)}=X$ $^{(f)}$,
immediately implies that%
\begin{equation}
\dsum\nolimits_{k}m_{k}(\Delta _{x})_{k}=0.  \label{SM}
\end{equation}%
Furthermore, given that $m\Delta _{x}=-mv\Delta _{t}=-p\Delta _{t}$ yields\ 
\begin{equation}
\dsum\nolimits_{k}p_{k}(\Delta _{t})_{k}=0,  \label{SP}
\end{equation}%
where $p_{k}$ is the momentum of a particle of type $k$. We will use the
relations (\ref{SM}) and (\ref{SP}) for consistency checks.

\subsection{Complex KdV multi-soliton scattering}

Let us now see how the above applies to the scattering of multi-solitons
that are solutions of the KdV equation for the complex field $%
u(x,t)=p(x,t)+iq(x,t)$ with $p(x,t)$, $q(x,t)\in \mathbb{R}$ 
\begin{equation}
u_{t}+6uu_{x}+u_{xxx}=0.  \label{KdV}
\end{equation}%
Separating (\ref{KdV}) into its real and imaginary part one may view it of
course as set of coupled equations for the real fields $p(x,t)$ and $q(x,t)$%
. In the limits $\left( pq\right) _{x}\rightarrow pq_{x}$ and $%
q_{xxx}\rightarrow 0$ they reduce to some well studied systems, the
Hirota-Satsuma \cite{hirota1981soliton} and Ito equations \cite%
{ito1982symmetries}, respectively. The total mass, momentum and energy
associated to the solution $u(x,t)$ are defined as%
\begin{equation}
m=\dint\nolimits_{-\infty }^{\infty }udx,~\qquad ~p=\dint\nolimits_{-\infty
}^{\infty }u^{2}dx,~\qquad ~E=\dint\nolimits_{-\infty }^{\infty }\left(
2u^{3}-u_{x}^{2}\right) dx,  \label{mpe}
\end{equation}%
respectively. See section \ref{realitycond} for a derivation of these
expressions. We have to establish that these quantities are real as they are
meant to be physical, i.e. observable.

\subsubsection{Properties of the one-soliton solutions}

First we need to compute complex solutions to the KdV equation. We recall
that they may be constructed for instance from Hirota's direct method \cite%
{Hirota}. Defining the quantities $\eta _{\mu ;\alpha }:=\alpha x-\alpha
^{3}t+\mu $, we consider the $\tau $-function for a one-soliton%
\begin{equation}
\tau _{\mu ;\alpha }(x,t)=1+e^{\eta _{\mu ;\alpha }},
\end{equation}%
from which the corresponding solution to the complex KdV equation is
obtained as $u(x,t)=2[\ln \tau (x,t)]_{xx}$. Taking the value for $\mu $ in
a form that respects the $\mathcal{PT}$-symmetry of the solutions, i.e.
purely imaginary $\mu =i\theta $ with $\theta \in \mathbb{R}$, we obtain 
\cite{CenFring}%
\begin{equation}
u_{i\theta ;\alpha }(x,t)=\frac{\alpha ^{2}+\alpha ^{2}\cos \theta \cosh
(\alpha x-\alpha ^{3}t)}{\left[ \cos \theta +\cosh (\alpha x-\alpha ^{3}t)%
\right] ^{2}}-i\frac{\alpha ^{2}\sin \theta \sinh (\alpha x-\alpha ^{3}t)}{%
\left[ \cos \theta +\cosh (\alpha x-\alpha ^{3}t)\right] ^{2}}.~~~~
\label{newS}
\end{equation}%
We may restrict ourselves to this choice, because $\mathcal{PT}$-symmetry
breaking choices, such as $\mu =\kappa +i\theta $ with $\kappa \in \mathbb{R}
$, can be converted to the form (\ref{newS}) by simple shifts in $x$ or $t$, 
$u_{\kappa +i\theta ;\alpha }(x,t)=u_{i\theta ;\alpha }(x+\kappa /\alpha
,t)=u_{i\theta ;\alpha }(x,t-\kappa /\alpha ^{3})$. The former may then be
absorbed in the limits of integrals when computing physical quantities and
the latter may be neglected when considering conserved charges. Using this
solution we compute the quantities as defined in (\ref{mpe}). For the mass
of the complex one-soliton we obtain always a real value 
\begin{equation}
m_{\alpha }=\dint\nolimits_{-\infty }^{\infty }u_{i\theta ;\alpha
}(x,t)dx=\left. \frac{\alpha \sinh (\alpha x-\alpha ^{3}t)+i\alpha \sin
\theta }{\cos \theta +\cosh (\alpha x-\alpha ^{3}t)}\right\vert _{x=-\infty
}^{\infty }=2\alpha .  \label{mass}
\end{equation}%
Thus only $p(x,t)$ component contributes to the mass of the soliton $u(x,t)$%
, so that $q(x,t)$ may be viewed as a massless soliton. Likewise, the
momentum of the one-soliton turns out to be always real%
\begin{eqnarray}
p_{\alpha } &=&\dint\nolimits_{-\infty }^{\infty }u_{i\theta ;\alpha
}^{2}dx=\left. \frac{\alpha ^{3}\sinh \eta _{0;\alpha }\left[ 5+6\cos \theta
\cosh \eta _{0;\alpha }+\cosh \left( 2\eta _{0;\alpha }\right) \right] }{%
6\left( \cos \theta +\cosh \eta _{0;\alpha }\right) ^{3}}\right\vert
_{x=-\infty }^{\infty }  \notag \\
&&\left. +~i\frac{\alpha ^{3}\sin \theta \left( 5+\cos (2\theta )+6\cos
\theta \cosh \left( \eta _{0;\alpha }\right) \right) }{6\left( \cos \theta
+\cosh \eta _{0;\alpha }\right) ^{3}}\right\vert _{x=-\infty }^{\infty }=%
\frac{2}{3}\alpha ^{3},  \label{impuls}
\end{eqnarray}%
and the value for the energy was reported in \cite{CenFring} to be real as
well%
\begin{equation}
E_{\alpha }=\dint\nolimits_{-\infty }^{\infty }\left[ 2u_{i\theta ;\alpha
}^{3}-\left( u_{i\theta ;\alpha }\right) _{x}^{2}\right] dx=\frac{2}{5}%
\alpha ^{5}.
\end{equation}%
In comparison with \cite{CenFring} we have rescaled the energy by a factor
of $-2$ for reason that become apparent in section 3.

We now follow \cite{fring1994vertex} by choosing a reference frame\ that
tracks a distinct point on the soliton, such the crest or trough of the
wave. Tracking the one-soliton solution by keeping the traveling wave
coordinate $x-\alpha ^{2}t$ at a fixed value, we obtain for the real part of
the solution (\ref{newS}) the constant values%
\begin{eqnarray}
p_{i\theta ;\alpha }\left[ t\alpha ^{2},t\right] &=&\frac{\alpha ^{2}}{2}%
\sec ^{2}\left( \frac{\theta }{2}\right) =:\hat{P}_{\alpha }(\theta ),\qquad
\label{one} \\
p_{i\theta ;\alpha }\left[ t\alpha ^{2}\pm \frac{1}{\alpha }\Delta
_{r}(\theta ),t\right] &=&-\frac{\alpha ^{2}}{4}\cot ^{2}\left( \theta
\right) =:\check{P}_{\alpha }(\theta ),  \label{two}
\end{eqnarray}%
corresponding to a maximum and two minima, respectively, with shift function%
\begin{equation}
\Delta _{r}(\theta ):=\func{arccosh}(\cos \theta -2\sec \theta ).
\label{delr}
\end{equation}%
Notice that the minima only emerge when $\theta \in \left( (4n+1)\pi
/2,(4n+3)\pi /2\right) $ with $n\in \mathbb{Z}$ as otherwise the argument of
the $\func{arccosh}$ in (\ref{delr}) is smaller 1. For the imaginary part we
define the shift function%
\begin{equation}
\Delta _{i}(\theta ):=\func{arccosh}\left[ \frac{1}{2}\cos \theta +\frac{%
\sqrt{2}}{4}\sqrt{17+\cos (2\theta )}\right] ,  \label{deli}
\end{equation}%
and compute the minimal and maximal values%
\begin{equation}
q_{i\theta ;\alpha }\left[ t\alpha ^{2}\pm \frac{1}{\alpha }\Delta
_{i}(\theta ),t\right] =\mp \frac{8\alpha ^{2}\sin \theta \sqrt{5+\cos
(2\theta )+\sqrt{2}\cos \theta \sqrt{17+\cos (2\theta )}}}{\left[ 6\cos
\theta +\sqrt{2}\sqrt{17+\cos (2\theta )}\right] ^{2}}=:\mp Q_{\alpha
}(\theta ).  \label{three}
\end{equation}%
With these shifts the real part of the one-soliton solution is simply fixed
to remain on the crest of the wave as time evolves in (\ref{one}) or on
either of the two minima in (\ref{two}) when they exist. For the imaginary
part we have the option to track either the crest or trough as specified in (%
\ref{three}).

We summarize these features in figure \ref{EinsS}.

\FIGURE{ \epsfig{file=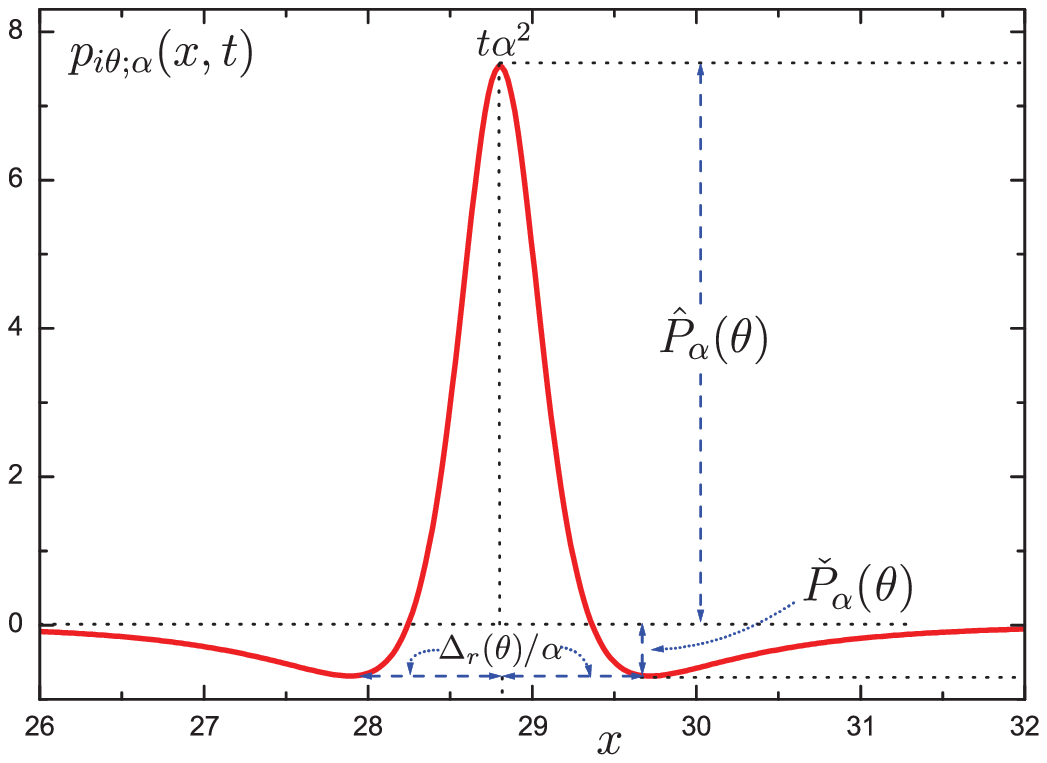,width=7.2cm}   \epsfig{file=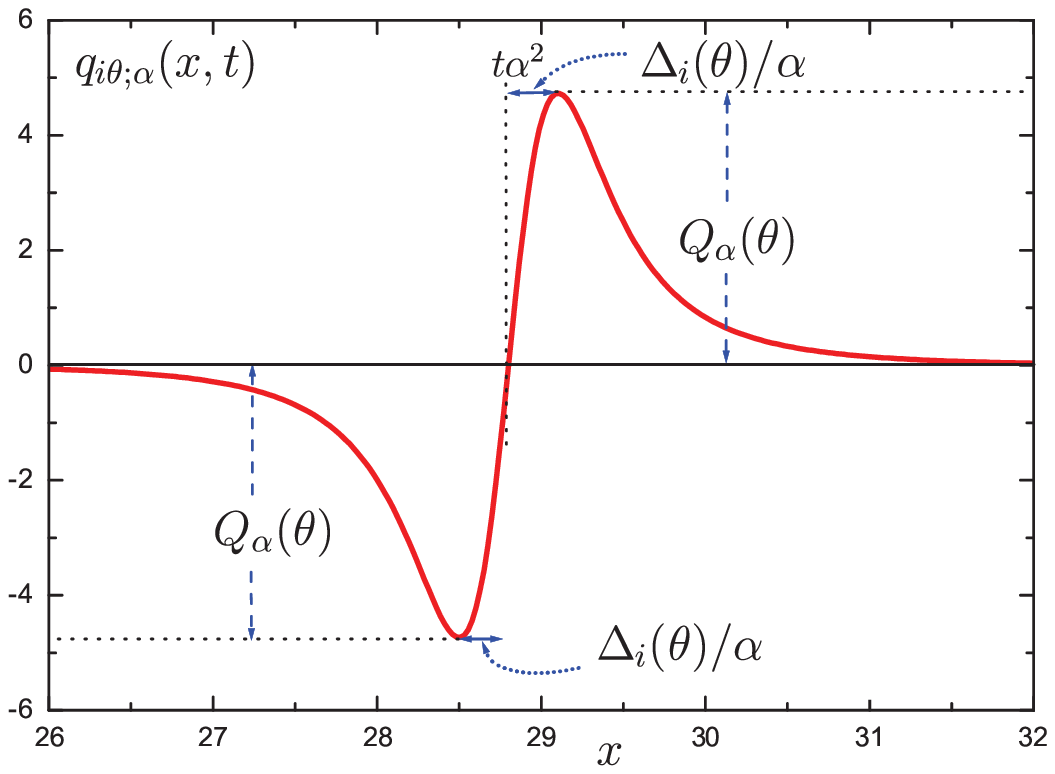,width=7.2cm}
\caption{$\mathcal{PT}$-symmetric one-soliton solution (\ref{newS}) of the KdV equation (\ref{KdV}) with $\alpha=6/5$ and $\theta=6/5 \pi $ at time $t=20$. }
        \label{EinsS}}

\subsubsection{Properties of nondegenerate two-soliton solutions}

Next we consider the complex two-soliton solution. Abbreviating the
reoccurring constant $\varkappa (\alpha ,\beta ):=$ $(\alpha -\beta
)^{2}/(\alpha +\beta )^{2}$, the two-soliton $\tau $-function is compactly
expressed as 
\begin{equation}
\tau _{\mu ,\nu ;\alpha ,\beta }(x,t)=1+e^{\eta _{\mu ;\alpha }}+e^{\eta
_{\nu ;\beta }}+\varkappa (\alpha ,\beta )e^{\eta _{\mu ;\alpha }+\eta _{\nu
;\beta }},
\end{equation}%
from which, by using again the transformation $u(x,t)=2[\ln \tau (x,t)]_{xx}$%
, we compute the solution 
\begin{equation}
\begin{array}{c}
u_{i\theta ,i\phi ;\alpha ,\beta }=\frac{2\left[ \alpha ^{2}e^{\eta
_{i\theta ;\alpha }}+\beta ^{2}e^{\eta _{i\phi ;\beta }}+\varkappa (\alpha
,\beta )\left( \alpha ^{2}e^{\eta _{i\theta ;\alpha }+2\eta _{i\phi ;\beta
}}+\beta ^{2}e^{2\eta _{i\theta ;\alpha }+\eta _{i\phi ;\beta }}\right)
+2(\alpha -\beta )^{2}e^{\eta _{i\theta ;\alpha }+\eta _{i\phi ;\beta }}%
\right] }{\tau _{i\theta ,i\phi ;\alpha ,\beta }^{2}}.%
\end{array}
\label{2S}
\end{equation}%
We now have to track the one-soliton contributions within the solution (\ref%
{2S}) and according to our definitions (\ref{DX}) and (\ref{DT}) we need to
compare the values in the infinite past with the one in the infinite future
in order to find the lateral displacements and time-delays. From (\ref{one})
we can read off which frame we have to choose. Tracking the maxima for the
real part of the two-soliton solution (\ref{2S}), we compute the asymptotic
values%
\begin{equation}
\begin{array}{l}
\lim\limits_{t\rightarrow -\infty }p_{i\theta ,i\phi ;\alpha ,\beta }\left[
t\alpha ^{2},t\right] =\lim\limits_{t\rightarrow +\infty }p_{i\theta ,i\phi
;\alpha ,\beta }\left[ t\alpha ^{2}+\delta _{\alpha }^{\alpha ,\beta },t%
\right] =\hat{P}_{\alpha }(\theta ), \\ 
\lim\limits_{t\rightarrow +\infty }p_{i\theta ,i\phi ;\alpha ,\beta }\left[
t\beta ^{2},t\right] =\lim\limits_{t\rightarrow -\infty }p_{i\theta ,i\phi
;\alpha ,\beta }\left[ t\beta ^{2}+\delta _{\beta }^{\alpha ,\beta },t\right]
=\hat{P}_{\beta }(\phi ),%
\end{array}
\label{2real}
\end{equation}%
where for definiteness we have taken the ordering $\alpha >\beta $ and
furthermore abbreviated the quantity%
\begin{equation}
\delta _{x}^{y,z}:=\frac{2}{x}\ln \left( \frac{y+z}{y-z}\right) ,
\end{equation}%
for conciseness. See appendix A for the details of this computation.
According to our definitions (\ref{DX}) and (\ref{DT}) we can now read off
the lateral displacements and time-delays from the asymptotic values in (\ref%
{2real}) by comparing the infinite future and the infinite past. For the
soliton with velocity $\alpha ^{2}$ we find 
\begin{equation}
(\Delta _{x})_{\alpha }=\delta _{\alpha }^{\alpha ,\beta },\qquad \text{%
and\qquad }(\Delta _{t})_{\alpha }=-\frac{1}{\alpha ^{2}}\delta _{\alpha
}^{\alpha ,\beta },  \label{dx}
\end{equation}%
and for the soliton with velocity $\beta ^{2}$ we identify%
\begin{equation}
(\Delta _{x})_{\beta }=-\delta _{\beta }^{\alpha ,\beta },\qquad \text{%
and\qquad }(\Delta _{t})_{\beta }=\frac{1}{\beta ^{2}}\delta _{\beta
}^{\alpha ,\beta }.  \label{dt}
\end{equation}

\FIGURE{  \epsfig{file=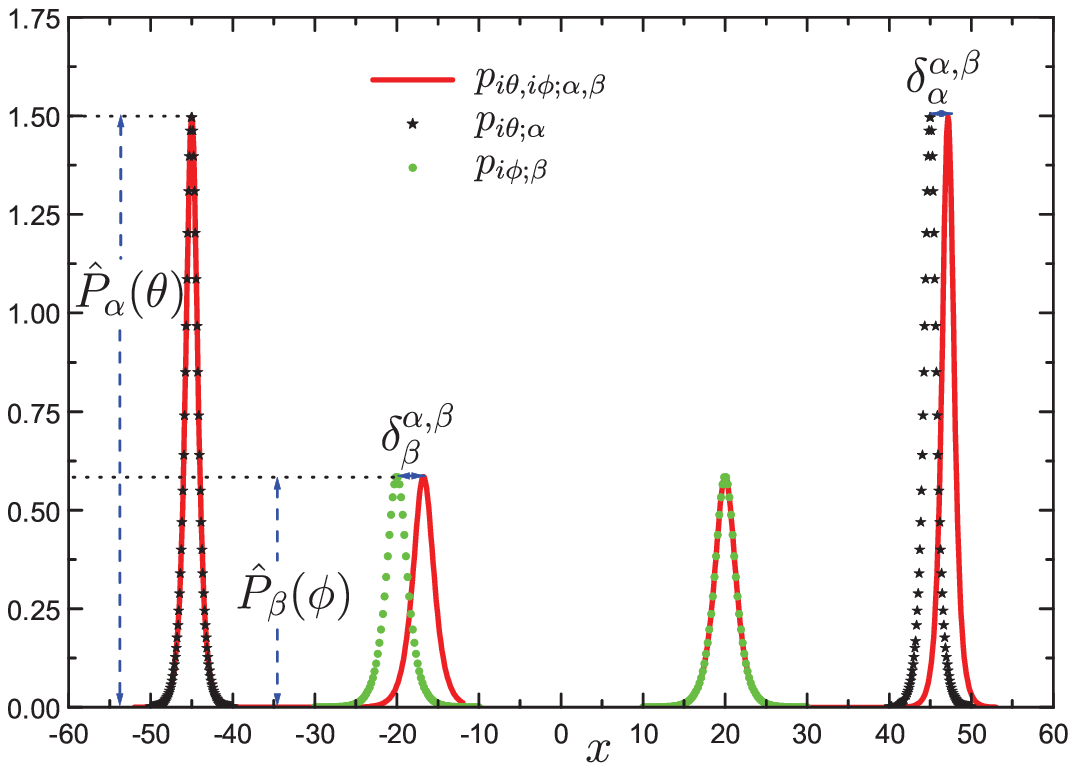,width=7.2cm}  \epsfig{file=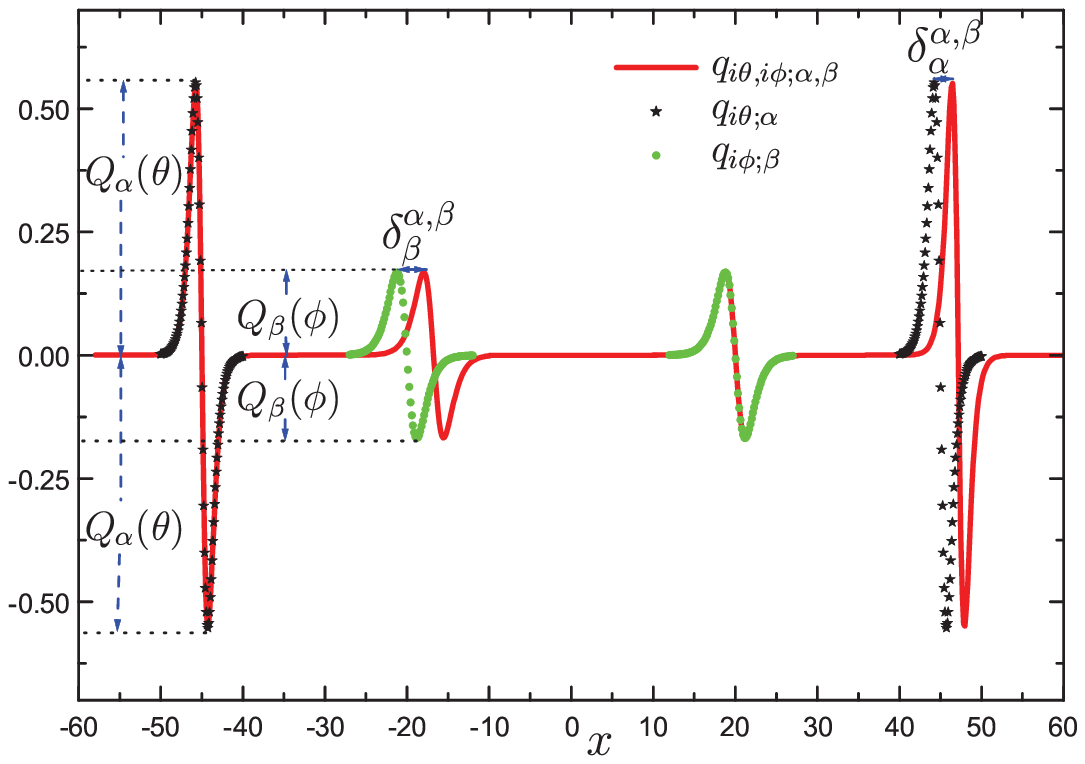,width=7.2cm} 
\caption{Lateral displacements for the complex $\mathcal{PT}$-symmetric two-soliton KdV solution (\ref{2S}) with $\alpha=3/2$, $\beta=1$, $\theta= \pi /3$ and $\phi =  \pi/4$. 
The plots in the negative and positive regime of $x$ correspond to the time taken to be $t=-20$ and $t=20$, respectively.}
        \label{Delay}}

Using the values for the masses and momenta computed in (\ref{mass})\ and (%
\ref{impuls}), we verify that the quantities (\ref{dx}) and (\ref{dt})
indeed satisfy the consistency relations (\ref{SM}) and (\ref{SP}), since%
\begin{eqnarray}
m_{\alpha }(\Delta _{x})_{\alpha }+m_{\beta }(\Delta _{x})_{\beta }
&=&2\alpha \delta _{\alpha }^{\alpha ,\beta }-2\beta \delta _{\beta
}^{\alpha ,\beta }=0, \\
p_{\alpha }(\Delta _{t})_{\alpha }+p_{\beta }(\Delta _{t})_{\beta } &=&-%
\frac{2}{3}\alpha ^{3}\frac{1}{\alpha ^{2}}\delta _{\alpha }^{\alpha ,\beta
}+\frac{2}{3}\beta ^{3}\frac{1}{\beta ^{2}}\delta _{\beta }^{\alpha ,\beta
}=0.
\end{eqnarray}
We may of course also track any other point and should obtain the same
values for the displacement and delays. For instance, tracking the minima
for the real part we compute 
\begin{equation}
\begin{array}{l}
\lim\limits_{t\rightarrow -\infty }p_{i\theta ,i\phi ;\alpha ,\beta }\left[
t\alpha ^{2}\pm \frac{\Delta _{r}(\theta )}{\alpha },t\right]
=\lim\limits_{t\rightarrow +\infty }p_{i\theta ,i\phi ;\alpha ,\beta }\left[
t\alpha ^{2}\pm \frac{\Delta _{r}(\theta )}{\alpha }+\delta _{\alpha
}^{\alpha ,\beta },t\right] =\check{P}_{\alpha }(\theta ), \\ 
\lim\limits_{t\rightarrow +\infty }p_{i\theta ,i\phi ;\alpha ,\beta }\left[
t\beta ^{2}\pm \frac{\Delta _{r}(\phi )}{\beta },t\right] =\lim\limits_{t%
\rightarrow -\infty }p_{i\theta ,i\phi ;\alpha ,\beta }\left[ t\beta ^{2}\pm 
\frac{\Delta _{r}(\phi )}{\beta }+\delta _{\beta }^{\alpha ,\beta },t\right]
=\check{P}_{\beta }(\phi ),%
\end{array}
\label{PH}
\end{equation}%
or tracking the minimum or maximum for the imaginary part we obtain%
\begin{equation}
\begin{array}{l}
\lim\limits_{t\rightarrow -\infty }q_{i\theta ,i\phi ;\alpha ,\beta }\left[
t\alpha ^{2}\pm \frac{\Delta _{i}(\theta )}{\alpha },t\right]
=\lim\limits_{t\rightarrow +\infty }q_{i\theta ,i\phi ;\alpha ,\beta }\left[
t\alpha ^{2}\pm \frac{\Delta _{i}(\theta )}{\alpha }+\delta _{\alpha
}^{\alpha ,\beta },t\right] =\mp Q_{\alpha }(\theta ), \\ 
\lim\limits_{t\rightarrow +\infty }q_{i\theta ,i\phi ;\alpha ,\beta }\left[
t\beta ^{2}\pm \frac{\Delta _{i}(\phi )}{\beta },t\right] =\lim\limits_{t%
\rightarrow -\infty }q_{i\theta ,i\phi ;\alpha ,\beta }\left[ t\beta ^{2}\pm 
\frac{\Delta _{i}(\phi )}{\beta }+\delta _{\beta }^{\alpha ,\beta },t\right]
=\mp Q_{\beta }(\phi ).%
\end{array}
\label{2im}
\end{equation}%
We depict these features in figure \ref{Delay}.

We also remark here that the time-delays are important to clarify the
precise relation between the solutions obtained from the Hirota method (\ref%
{2S}) and those constructed via B\"{a}cklund transformations \cite{CenFring}%
\begin{equation}
u_{\mu ,\nu ;\alpha ,\beta }^{B}(x,t)=\frac{\alpha ^{2}-\beta ^{2}}{2}\frac{%
\beta ^{2}\func{sech}\left[ \frac{1}{2}(\beta x-\beta ^{3}t+\nu )\right]
^{2}-\alpha ^{2}\func{sech}\left[ \frac{1}{2}(\alpha x-\alpha ^{3}t+\mu )%
\right] ^{2}}{\left[ \alpha \tanh \left[ \frac{1}{2}(\alpha x-\alpha
^{3}t+\mu )\right] -\beta \tanh \left[ \frac{1}{2}(\beta x-\beta ^{3}t+\nu )%
\right] \right] ^{2}}.  \label{twos}
\end{equation}%
Tracking the maxima of the real part of (\ref{twos}) we compute the
asymptotic values%
\begin{equation}
\begin{array}{l}
\lim\limits_{t\rightarrow +\infty }p_{i\theta ,i\phi ;\alpha ,\beta }^{B} 
\left[ \alpha ^{2}t+\frac{\delta _{\alpha }^{\alpha ,\beta }}{2}\right]
=\lim\limits_{t\rightarrow -\infty }p_{i\theta ,i\phi ;\alpha ,\beta }^{B}%
\left[ \alpha ^{2}t-\frac{\delta _{\alpha }^{\alpha ,\beta }}{2}\right] =%
\hat{P}_{\alpha }(\theta +\pi ) \\ 
\lim\limits_{t\rightarrow +\infty }p_{i\theta ,i\phi ;\alpha ,\beta }^{B} 
\left[ \beta ^{2}t-\frac{\delta _{\beta }^{\alpha ,\beta }}{2}\right]
=\lim\limits_{t\rightarrow -\infty }p_{i\theta ,i\phi ;\alpha ,\beta }^{B}%
\left[ \beta ^{2}t+\frac{\delta _{\beta }^{\alpha ,\beta }}{2}\right] =\hat{P%
}_{\beta }(\phi )%
\end{array}
\label{PB}
\end{equation}

\FIGURE{ \epsfig{file=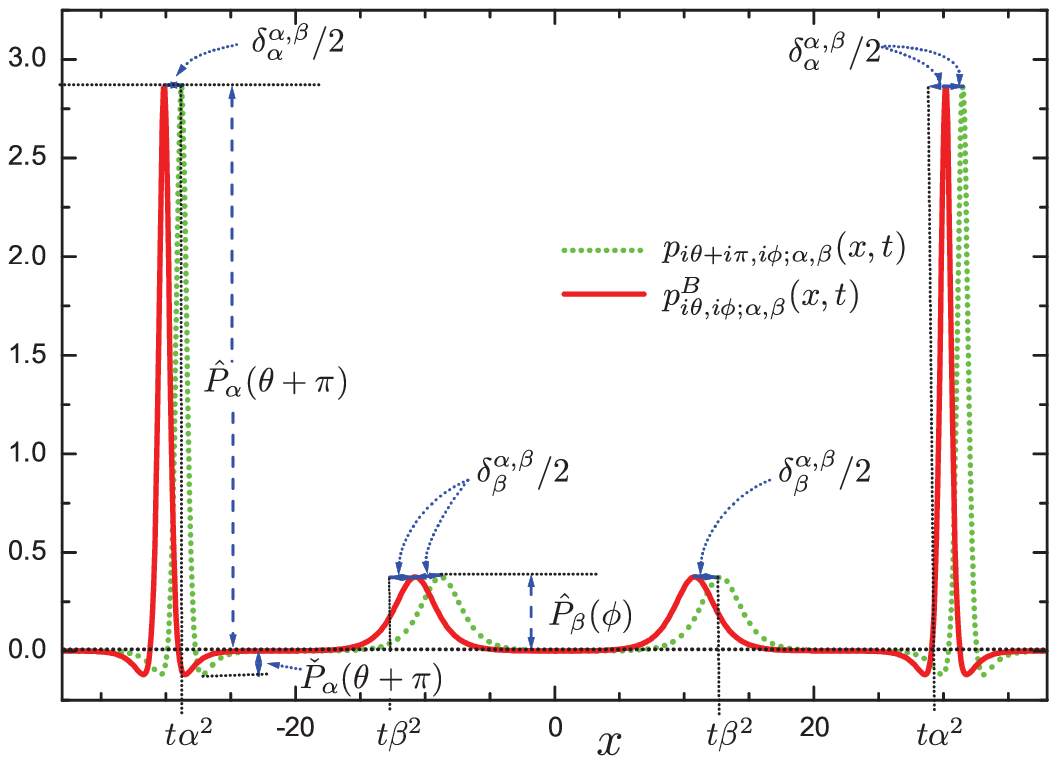,width=7.2cm}  \epsfig{file=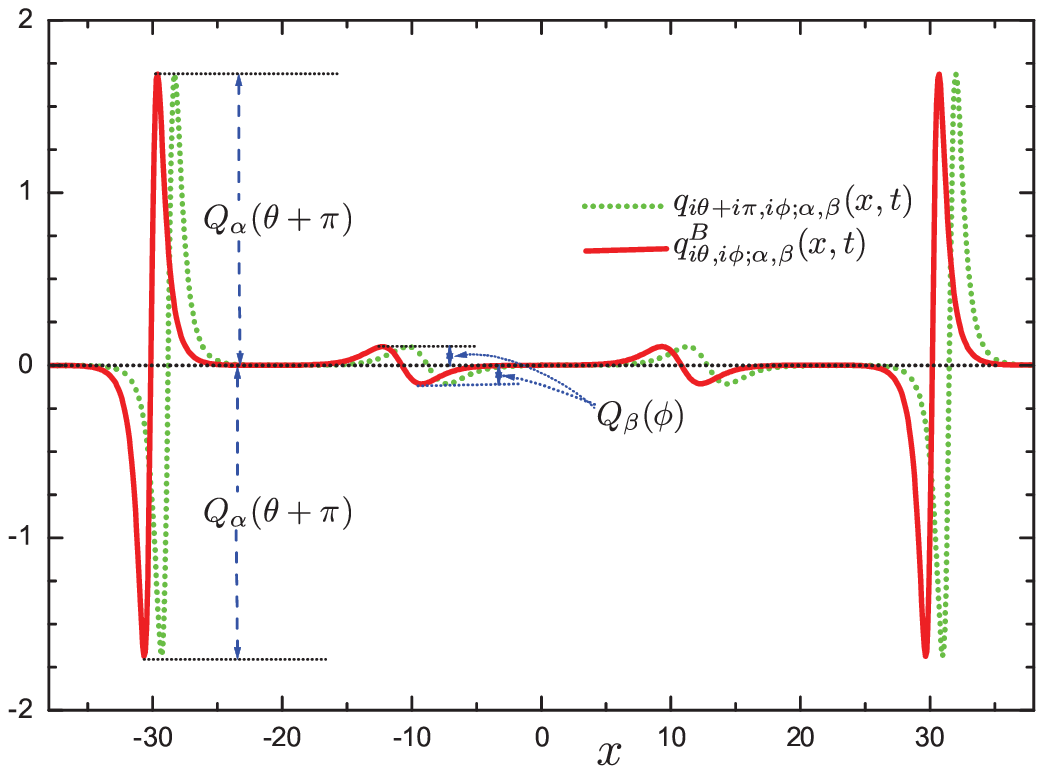,width=7.2cm} 
\caption{Complex Hirota two-soliton KdV solution (\ref{2S}) versus two-soliton KdV solution obtained from B\"{a}cklund transformations (\ref{twos}) for $\alpha=1.2$, $\beta=0.8$, $\theta= \pi/3$ and $\phi= \pi/4$. 
The plots in the negative and positive regime of $x$ correspond to the time taken to be $t=-20$ and $t=20$, respectively.}
        \label{HirBack}}

Comparing the real parts of the one-soliton contribution within the B\"{a}%
cklund two-soliton solution with those in the Hirota solution, we find that
the slower and faster one are delayed by the amount $\delta _{\beta
}^{\alpha ,\beta }/2$ and $\delta _{\alpha }^{\alpha ,\beta }/2$,
respectively. Thus the two types of solutions are not simply shifted by an
overall amount, but each of the individual one-soliton contributions shifted
by a different amount relative to each other. Furthermore, comparing (\ref%
{PH}) and (\ref{PB}) we observe that we require a shift in $\theta $ by $\pi 
$ in the Hirota solution in order to obtain the same qualitative features in
both solution, in the sense of matching amplitudes and occurrence of minima.
Overall we obtain the same total delays (\ref{dx}) from any of the solutions.

Tracking the minima and maxima for the imaginary part we obtain 
\begin{equation}
\begin{array}{l}
\!\!\lim\limits_{t\rightarrow +\infty }q_{i\theta ,i\phi ;\alpha ,\beta
}^{B} \left[ \alpha ^{2}t+\frac{\delta _{\alpha }^{\alpha ,\beta }}{2}\pm 
\frac{\Delta _{i}(\theta +\pi )}{\alpha }\right] =\!\!\lim\limits_{t%
\rightarrow -\infty }q_{i\theta ,i\phi ;\alpha ,\beta }^{B}\left[ \alpha
^{2}t-\frac{\delta _{\alpha }^{\alpha ,\beta }}{2}\pm \frac{\Delta
_{i}(\theta +\pi )}{\alpha }\right] =\mp Q_{\alpha }(\theta +\pi ) \\ 
\!\!\lim\limits_{t\rightarrow +\infty }q_{i\theta ,i\phi ;\alpha ,\beta
}^{B} \left[ \beta ^{2}t-\frac{\delta _{\beta }^{\alpha ,\beta }}{2}\pm 
\frac{1}{\beta }\Delta _{i}(\phi )\right] =\!\!\lim\limits_{t\rightarrow
-\infty }q_{i\theta ,i\phi ;\alpha ,\beta }^{B}\left[ \beta ^{2}t+\frac{%
\delta _{\beta }^{\alpha ,\beta }}{2}\pm \frac{1}{\beta }\Delta _{i}(\phi )%
\right] =\mp Q_{\beta }(\phi ).%
\end{array}%
\end{equation}

As expected, we observe from this that the imaginary parts of the soliton
solutions are displaced by the same amount as the real parts, but for the
faster soliton we also acquired an overall minus sign.

These features are displayed in figure \ref{HirBack}.

We notice that the time-delay caused by the scattering between the faster
and the slower soliton is the same in both solutions, i.e. they are
preserved quantities.

\subsubsection{Properties of degenerate two-soliton solutions}

\label{sectwodeg}

As discussed in \cite{FranciscoAndreas}, the degenerate solutions for which
some of the energies are the same are quite special. In general the limit $%
\beta \rightarrow \alpha $ to degenerate energies in a multi-soliton
solution is divergent. However, as discussed in detail in \cite%
{FranciscoAndreas}, in a complex setting, when tuning certain parameters it
can be performed consistently, leading to the solution%
\begin{equation}
u_{i\theta ,i\phi ;\alpha ,\alpha }(x,t)=\frac{2\alpha ^{2}\left[ \left(
\alpha x-3\alpha ^{3}t+i\phi \right) \sinh \left( \eta _{i\theta ;\alpha
}\right) -2\cosh \left( \eta _{i\theta ;\alpha }\right) -2\right] }{\left[
\alpha x-3\alpha ^{3}t+i\phi +\sinh \left( \eta _{i\theta ;\alpha }\right) %
\right] ^{2}}.  \label{utwo}
\end{equation}%
Defining the time-dependent displacement%
\begin{equation}
\Delta (t):=\frac{1}{\alpha }\ln \left( 4\alpha ^{3}\left\vert t\right\vert
\right) ,  \label{delt}
\end{equation}%
we track the maximum for the real part of the degenerate solution (\ref{utwo}%
) as%
\begin{eqnarray}
\lim_{t\rightarrow +\infty }p_{i\theta ,i\phi ;\alpha ,\alpha }\left[
t\alpha ^{2}-\Delta (t),t\right] &=&\lim_{t\rightarrow -\infty }p_{i\theta
,i\phi ;\alpha ,\alpha }\left[ t\alpha ^{2}+\Delta (t),t\right] =\hat{P}%
_{\alpha }(\theta ),  \label{s1} \\
\lim_{t\rightarrow +\infty }p_{i\theta ,i\phi ;\alpha ,\alpha }\left[
t\alpha ^{2}+\Delta (t),t\right] &=&\lim_{t\rightarrow -\infty }p_{i\theta
,i\phi ;\alpha ,\alpha }\left[ t\alpha ^{2}-\Delta (t),t\right] =\hat{P}%
_{\alpha }(\theta +\pi ),  \label{s2}
\end{eqnarray}%
See appendix A for a derivation of these asymptotic expressions. Comparing (%
\ref{s1}) with (\ref{one}) we observe that the time-dependent shift is
tracking the maximum of the one-soliton solution. The second maximum (\ref%
{s2}) corresponds to the one-soliton solution (\ref{one}) with $\theta
\rightarrow \theta +\pi $, which relates the $\func{sech}^{2}$ to the $\func{%
csch}^{2}$ solution. We expect these solutions to emerge as in the real case
they correspond to the two independent solutions from which the degenerate
one (\ref{utwo}) was constructed in \cite{FranciscoAndreas} when using
Wronskians involving Jordan states. Moreover, regarding the internal
structure of the degenerate two-soliton we deduce that the one-solitons with
amplitude $\hat{P}_{\alpha }(\theta )$ and $\hat{P}_{\alpha }(\theta +\pi )$
are laterally displaced by $-2\Delta (t)$ and $2\Delta (t)$, respectively,
as a result of the scattering process. When $t\rightarrow \pm \infty $ the
displacements tend to infinity as we somehow expect from the displacement (%
\ref{dx}) which diverges when $\beta \rightarrow \alpha $.

As we have seen, the real part of one of the one-soliton solutions also
develops two minima, for which we compute the limits%
\begin{equation}
\lim_{t\rightarrow \sigma \infty }p_{i\theta ,i\phi ;\alpha ,\alpha }\left[
t\alpha ^{2}-\sigma \Delta (t)\pm \frac{1}{\alpha }\Delta _{r}(\theta ),t%
\right] =\check{P}_{\alpha }(\theta ).
\end{equation}%
where $\sigma $ can be $+1$ or $-1$. Since the two one-solitons are
relatively shifted to each other by $\theta \rightarrow \theta +\pi $ it
follows from the remarks after (\ref{delr}) that these minima can only
emerge in one of the two solitons.

For the imaginary part we compute the eight limits%
\begin{eqnarray}
\lim\limits_{t\rightarrow \sigma \infty }q_{i\theta ,i\phi ;\alpha ,\alpha }
\left[ t\alpha ^{2}-\sigma \Delta (t)\pm \frac{1}{\alpha }\Delta _{i}(\theta
),t\right]  &=&\mp Q_{\alpha }(\theta ), \\
\lim\limits_{t\rightarrow \sigma \infty }q_{i\theta ,i\phi ;\alpha ,\alpha }
\left[ t\alpha ^{2}+\sigma \Delta (t)\pm \frac{1}{\alpha }\Delta _{i}(\theta
+\pi ),t\right]  &=&\pm Q_{\alpha }(\theta +\pi ).
\end{eqnarray}%
We also observe from the imaginary part that the overall time-delays are $%
\pm 2\Delta (t)$.

\subsubsection{Properties of nondegenerate three-soliton solutions}

Let us now consider the three-soliton solution for which the $\tau $%
-function reads 
\begin{eqnarray}
\tau _{\mu ,\nu ,\rho ;\alpha ,\beta ,\gamma }(x,t) &=&1+e^{\eta _{\mu
;\alpha }}+e^{\eta _{\nu ;\beta }}+e^{\eta _{\rho ;\gamma }}+\varkappa
(\alpha ,\beta )e^{\eta _{\mu ;\alpha }+\eta _{\nu ;\beta }}+\varkappa
(\alpha ,\gamma )e^{\eta _{\mu ;\alpha }+\eta _{\rho ;\gamma }}~~~ \\
&&+\varkappa (\beta ,\gamma )e^{\eta _{\nu ;\beta }+\eta _{\rho ;\gamma
}}+\varkappa (\alpha ,\beta )\varkappa (\alpha ,\gamma )\varkappa (\beta
,\gamma )e^{\eta _{\mu ;\alpha }+\eta _{\nu ;\beta }+\eta _{\rho ;\gamma }},
\notag
\end{eqnarray}%
leading to the three-soliton solution%
\begin{equation}
u_{i\theta ,i\phi ,i\vartheta ;\alpha ,\beta ,\gamma }(x,t)=2\left[ \ln \tau
_{i\theta ,i\phi ,i\vartheta ;\alpha ,\beta ,\gamma }(x,t)\right] _{xx},
\end{equation}%
which we do not report here explicitly. Assuming the ordering $\alpha >\beta
>\gamma $ we track the maxima for the real parts and compute the asymptotic
values 
\begin{equation}
\begin{array}{r}
\lim\limits_{t\rightarrow -\infty }p_{i\theta ,i\phi ,i\vartheta ;\alpha
,\beta ,\gamma }\left[ t\alpha ^{2},t\right] =\lim\limits_{t\rightarrow
+\infty }p_{i\theta ,i\phi ,i\vartheta ;\alpha ,\beta ,\gamma }\left[
t\alpha ^{2}+\delta _{\alpha }^{\alpha ,\beta }+\delta _{\alpha }^{\alpha
,\gamma },t\right] =\hat{P}_{\alpha }(\theta ), \\ 
\lim\limits_{t\rightarrow -\infty }p_{i\theta ,i\phi ,i\vartheta ;\alpha
,\beta ,\gamma }\left[ t\beta ^{2}+\delta _{\beta }^{\alpha ,\beta },t\right]
=\lim\limits_{t\rightarrow +\infty }p_{i\theta ,i\phi ,i\vartheta ;\alpha
,\beta ,\gamma }\left[ t\beta ^{2}+\delta _{\beta }^{\beta ,\gamma },t\right]
=\hat{P}_{\beta }(\phi ), \\ 
\lim\limits_{t\rightarrow -\infty }p_{i\theta ,i\phi ,i\vartheta ;\alpha
,\beta ,\gamma }\left[ t\gamma ^{2}+\delta _{\gamma }^{\alpha ,\gamma
}+\delta _{\gamma }^{\beta ,\gamma },t\right] =\lim\limits_{t\rightarrow
+\infty }p_{i\theta ,i\phi ,i\vartheta ;\alpha ,\beta ,\gamma }\left[
t\gamma ^{2},t\right] =\hat{P}_{\gamma }(\vartheta ).%
\end{array}%
\end{equation}%
When tracking the minima or maxima in the imaginary part we obtain 
\begin{equation*}
\begin{array}{r}
\!\!\!\!\lim\limits_{t\rightarrow -\infty }\!\!q_{i\theta ,i\phi ,i\vartheta
;\alpha ,\beta ,\gamma }\left[ t\alpha ^{2}\pm \frac{\Delta _{i}(\theta )}{%
\alpha },t\right] =\!\!\lim\limits_{t\rightarrow \infty }\!\!q_{i\theta
,i\phi ,i\vartheta ;\alpha ,\beta ,\gamma }\left[ t\alpha ^{2}+\delta
_{\alpha }^{\alpha ,\beta }+\delta _{\alpha }^{\alpha ,\gamma }\pm \frac{%
\Delta _{i}(\theta )}{\alpha },t\right] =\mp Q_{\alpha }(\theta ), \\ 
\!\!\!\!\lim\limits_{t\rightarrow -\infty }\!\!q_{i\theta ,i\phi ,i\vartheta
;\alpha ,\beta ,\gamma }\left[ t\beta ^{2}+\delta _{\beta }^{\alpha ,\beta
}\pm \frac{\Delta _{i}(\phi )}{\beta },t\right] =\!\!\lim\limits_{t%
\rightarrow \infty }\!\!q_{i\theta ,i\phi ,i\vartheta ;\alpha ,\beta ,\gamma
}\left[ t\beta ^{2}+\delta _{\beta }^{\beta ,\gamma }\pm \frac{\Delta
_{i}(\phi )}{\beta },t\right] =\mp Q_{\beta }(\phi ), \\ 
\!\!\!\!\lim\limits_{t\rightarrow -\infty }\!\!q_{i\theta ,i\phi ,i\vartheta
;\alpha ,\beta ,\gamma }\left[ t\gamma ^{2}+\delta _{\gamma }^{\alpha
,\gamma }+\delta _{\gamma }^{\beta ,\gamma }\pm \frac{\Delta _{i}(\vartheta )%
}{\gamma },t\right] =\!\!\lim\limits_{t\rightarrow \infty }\!\!q_{i\theta
,i\phi ,i\vartheta ;\alpha ,\beta ,\gamma }\left[ t\gamma ^{2}\pm \frac{%
\Delta _{i}(\vartheta )}{\gamma },t\right] =\mp Q_{\gamma }(\vartheta ).%
\end{array}%
\end{equation*}

Similarly as in the previous section for the two-soliton solution we read
off the lateral displacements form these expressions as 
\begin{equation}
(\Delta _{x})_{\alpha }=\delta _{\alpha }^{\alpha ,\beta }+\delta _{\alpha
}^{\alpha ,\gamma },\qquad (\Delta _{x})_{\beta }=\delta _{\beta }^{\beta
,\gamma }-\delta _{\beta }^{\alpha ,\beta },\text{\qquad }(\Delta
_{x})_{\gamma }=-\delta _{\gamma }^{\alpha ,\gamma }-\delta _{\gamma
}^{\beta ,\gamma }.
\end{equation}%
The corresponding time-delays are%
\begin{equation}
(\Delta _{t})_{\alpha }=-\frac{1}{\alpha ^{2}}\left( \delta _{\alpha
}^{\alpha ,\beta }+\delta _{\alpha }^{\alpha ,\gamma }\right) ,~~(\Delta
_{t})_{\beta }=\frac{1}{\beta ^{2}}\left( \delta _{\beta }^{\alpha ,\beta
}-\delta _{\beta }^{\beta ,\gamma }\right) ,\text{~~}(\Delta _{t})_{\gamma }=%
\frac{1}{\gamma ^{2}}\left( \delta _{\gamma }^{\alpha ,\gamma }+\delta
_{\gamma }^{\beta ,\gamma }\right) .  \label{del}
\end{equation}%
Once again we may use the values for the soliton mass (\ref{mass})\ and
momentum (\ref{impuls}) to confirm that these quantities satisfy the
consistency relation (\ref{SM}) and (\ref{SP}). As the values in (\ref{del})
are simply the sums of the scattering of two solitons, this confirms the
well known factorization property in integrable systems stating that a
multiple scattering process can always be understood as consecutive
scattering of two particles for which any ordering is equivalent \cite{Holl}%
. In the quantized version of the model this property is reflected in the
Yang-Baxter and bootstrap equations.

\subsubsection{Properties of degenerate three-soliton solutions}

\label{secthreedeg}

Let us now consider the degenerate three-soliton solution for which the
limit $\beta ,\gamma \rightarrow \alpha $ is carried out. In \cite%
{FranciscoAndreas} a solution to this scenario was reported as%
\begin{equation}
u_{i\theta ,i\phi ,i\vartheta ;\alpha ,\alpha ,\alpha }(x,t)=2\left\{ \ln 
\left[ \left( 1+\left( \eta _{i\vartheta ;\alpha }^{(3)}\right) ^{2}+\cosh
\eta _{i\theta ;\alpha }^{(1)}\right) \sinh \frac{\eta _{i\theta ;\alpha
}^{(1)}}{2}-\eta _{i\phi ;\alpha }^{(9)}\cosh \frac{\eta _{i\theta ;\alpha
}^{(1)}}{2}\right] \right\} _{xx},  \label{uthree}
\end{equation}%
with $\eta _{\mu ;\alpha }^{(\lambda )}:=\alpha x-\lambda \alpha ^{3}t+\mu $%
. In this case we find the three maxima for the real parts in the limits%
\begin{eqnarray}
\lim_{t\rightarrow \pm \infty }p_{i\theta ,i\phi ,i\vartheta ;\alpha ,\alpha
,\alpha }\left[ t\alpha ^{2},t\right] &=&\hat{P}_{\alpha }(\theta +\pi ),
\label{l1} \\
\lim_{t\rightarrow \pm \infty }p_{i\theta ,i\phi ,i\vartheta ;\alpha ,\alpha
,\alpha }\left[ t\alpha ^{2}\pm \bar{\Delta}(t),t\right] &=&\hat{P}_{\alpha
}(\theta ),  \label{l2}
\end{eqnarray}%
where the time-dependent displacement is defined as%
\begin{equation}
\bar{\Delta}(t):=\frac{1}{\alpha }\ln \left( 8\alpha ^{6}t^{2}\right) .
\label{delt2}
\end{equation}%
Thus we find the center soliton converging to the $\func{sech}^{2}$%
-one-soliton solution and the two outer ones to the $\func{csch}^{2}$%
-one-soliton solution. The outer ones keep moving away from the center as $%
\left\vert t\right\vert $ increases. Similarly as for the degenerate
two-soliton, the two individual outer solitons with amplitudes $\hat{P}%
_{\alpha }(\theta )$ are time-dependently displaced by the different amounts 
$\pm 2\bar{\Delta}(t)$. This might is not be obvious from the limits (\ref%
{l2}) as these one-solitons are identical, but they have actually exchanged
their position. The one-soliton in the center with amplitude $\hat{P}%
_{\alpha }(\theta +\pi )$ is not displaced or time-delayed and simply
travels identically to a one-soliton solution. The real part of the
solutions posses also minima in the regimes for $\theta $ as specified after
(\ref{delr}). For those we compute%
\begin{eqnarray}
\lim_{t\rightarrow \pm \infty }p_{i\theta ,i\phi ,i\vartheta ;\alpha ,\alpha
,\alpha }\left[ t\alpha ^{2}\pm \frac{1}{\alpha }\Delta _{r}(\theta +\pi ),t%
\right] &=&\check{P}_{\alpha }(\theta ), \\
\lim_{t\rightarrow \pm \infty }p_{i\theta ,i\phi ,i\vartheta ;\alpha ,\alpha
,\alpha }\left[ t\alpha ^{2}\pm \bar{\Delta}(t)\pm \frac{1}{\alpha }\Delta
_{r}(\theta ),t\right] &=&\check{P}_{\alpha }(\theta ).
\end{eqnarray}%
For the imaginary parts we evaluate%
\begin{eqnarray}
\lim_{t\rightarrow \sigma \infty }q_{i\theta ,i\phi ,i\vartheta ;\alpha
,\alpha ,\alpha }\left[ t\alpha ^{2}\pm \frac{1}{\alpha }\Delta _{i}(\theta
+\pi ),t\right] &=&\mp Q_{\alpha }(\theta +\pi ), \\
\lim_{t\rightarrow \sigma \infty }q_{i\theta ,i\phi ,i\vartheta ;\alpha
,\alpha ,\alpha }\left[ t\alpha ^{2}+\bar{\Delta}(t)\pm \frac{1}{\alpha }%
\Delta _{i}(\theta ),t\right] &=&\mp Q_{\alpha }(\theta ), \\
\lim_{t\rightarrow \sigma \infty }q_{i\theta ,i\phi ,i\vartheta ;\alpha
,\alpha ,\alpha }\left[ t\alpha ^{2}-\bar{\Delta}(t)\pm \frac{1}{\alpha }%
\Delta _{i}(\theta ),t\right] &=&\mp Q_{\alpha }(\theta ).
\end{eqnarray}%
Using these limits we deduce the same values for the displacements as from
the real part.

Next we consider the degenerate three-soliton solution for which only the
limit $\beta \rightarrow \alpha $ is carried out, such that only two of the
contributions are degenerate. We recall a solution for this from \cite%
{FranciscoAndreas} 
\begin{eqnarray}
u_{i\theta ,i\phi ,i\vartheta ;\alpha ,\alpha ,\gamma }(x,t) &=&2\left\{ \ln 
\left[ \cosh \frac{\eta _{i\vartheta ;\gamma }^{(1)}}{2}\left[ \frac{\alpha
^{2}+\gamma ^{2}}{8}\sinh \left( \eta _{i\theta ;\alpha }^{(1)}\right) -%
\frac{\alpha ^{2}-\gamma ^{2}}{8}\eta _{i\phi ;\alpha }^{(3)}\right] \right.
\right. \\
&&\left. \left. -\frac{\alpha \gamma }{2}\cosh ^{2}\left( \frac{\eta
_{i\theta ;\alpha }^{(1)}}{2}\right) \sinh \left( \frac{\eta _{i\vartheta
;\gamma }^{(1)}}{2}\right) \right] \right\} _{xx}.  \notag
\end{eqnarray}%
For the degenerated compound soliton the two maxima of the real part have
the properties%
\begin{eqnarray}
\lim_{t\rightarrow \sigma \infty }p_{i\theta ,i\phi ,i\vartheta ;\alpha
,\alpha ,\gamma }\left[ t\alpha ^{2}+\sigma \Delta (t)+\sigma \frac{\delta
_{\alpha }^{\alpha ,\gamma }}{2},t\right] &=&\hat{P}_{\alpha }(\theta ),
\label{31} \\
\lim_{t\rightarrow \sigma \infty }p_{i\theta ,i\phi ,i\vartheta ;\alpha
,\alpha ,\gamma }\left[ t\alpha ^{2}-\sigma \Delta (t)+\sigma \frac{\delta
_{\alpha }^{\alpha ,\gamma }}{2},t\right] &=&\hat{P}_{\alpha }(\theta +\pi ),
\end{eqnarray}%
and for the non-degenerated one-soliton contribution we compute%
\begin{equation}
\lim_{t\rightarrow \sigma \infty }p_{i\theta ,i\phi ,i\vartheta ;\alpha
,\alpha ,\gamma }\left[ t\gamma ^{2}-\sigma \delta _{\gamma }^{\alpha
,\gamma },t\right] =\hat{P}_{\gamma }(\vartheta ).  \label{11}
\end{equation}

The degenerate one-solitons with amplitudes $\hat{P}_{\alpha }(\theta )$ and 
$\hat{P}_{\alpha }(\theta +\pi )$ are now time-dependently displaced due to
the scattering amongst each other and in addition displaced by a constant
due to the scattering by%
\begin{equation}
(\Delta _{x})_{\alpha }^{\theta }=2\Delta (t)+\delta _{\alpha }^{\alpha
,\gamma }\qquad \text{and\qquad }(\Delta _{x})_{\alpha }^{\theta +\pi
}=-2\Delta (t)+\delta _{\alpha }^{\alpha ,\gamma }
\end{equation}
respectively. From (\ref{11}) we deduce%
\begin{equation}
(\Delta _{x})_{\gamma }=-2\delta _{\gamma }^{\alpha ,\gamma }
\end{equation}%
the constant displacement for the one-soliton with amplitude $\hat{P}%
_{\gamma }(\vartheta )$. Our consistency equation (\ref{SM}) is satisfied as%
\begin{equation}
m_{\alpha }(\Delta _{x})_{\alpha }^{\theta }+m_{\alpha }(\Delta
_{x})_{\alpha }^{\theta +\pi }+m_{\gamma }(\Delta _{x})_{\gamma }=4\alpha
\delta _{\alpha }^{\alpha ,\gamma }-4\gamma \delta _{\gamma }^{\alpha
,\gamma }=0.
\end{equation}%
We can argue similarly for the time-delays. Thus while the two individual
degenerate contributions are time-dependently displaced, there are in
addition nonvanishing constant contributions as a result of the scattering
with the remaining non-degenerate one-soliton. In a similar fashion as
above, these features are confirmed when tracking the minima in the real
part or the minima and maxima in the imaginary part.

These features are summarized in figure \ref{ThreeDelay}.

\FIGURE{  \epsfig{file=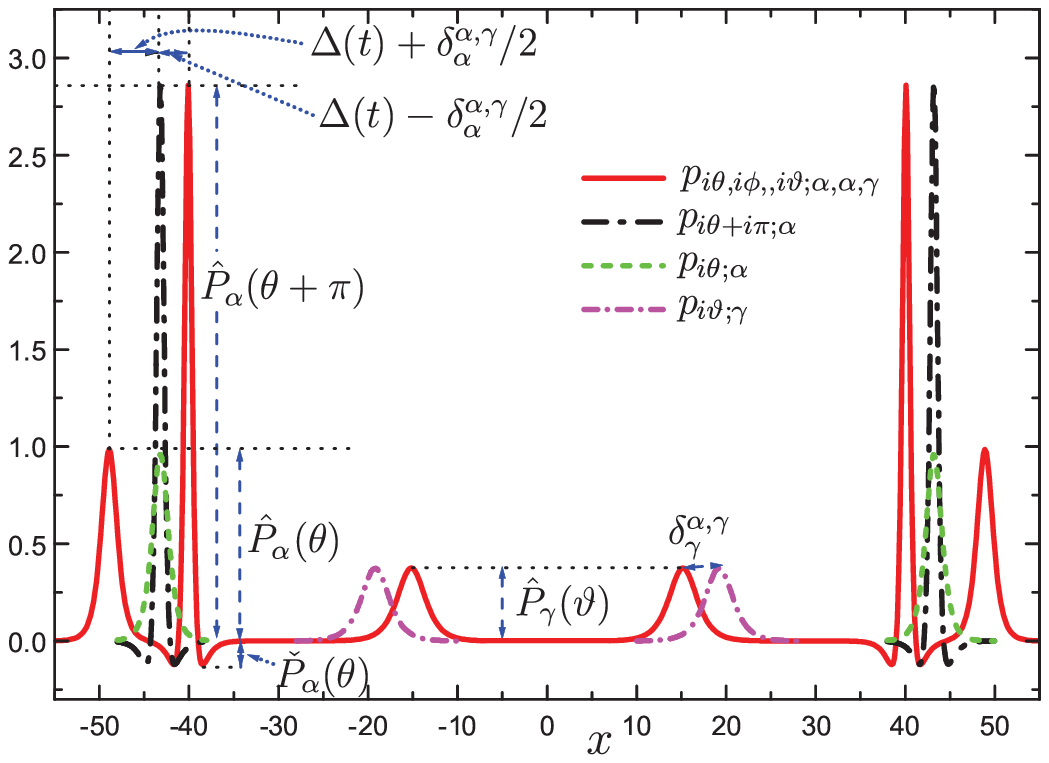,width=7.2cm}  \epsfig{file=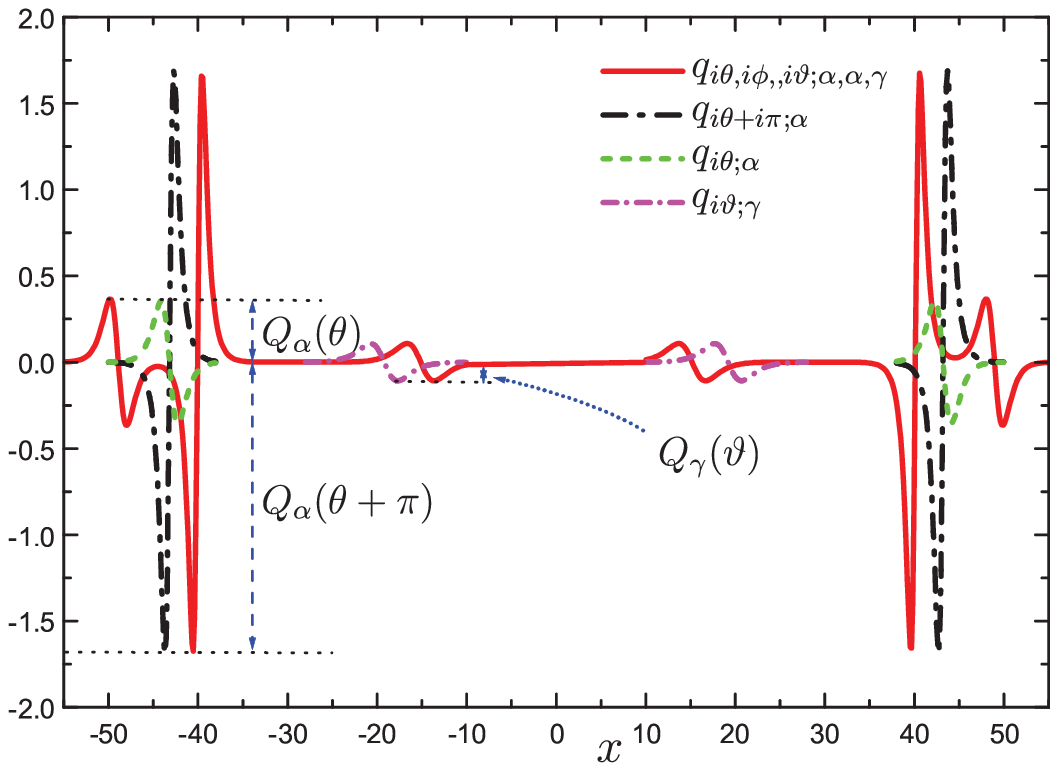,width=7.2cm} 
\caption{Time-delays for a complex $\mathcal{PT}$-symmetric three-soliton KdV solution with a compound two-soliton with $\alpha=6/5$, $\gamma=4/5$, $\theta=  \pi /3$ and $\vartheta=\phi =  \pi/4$. 
The plots in the negative and positive regime of $x$ correspond to the time taken to be $t=-30$ and $t=30$, respectively.}
        \label{ThreeDelay}}

\subsubsection{Properties of degenerate multi-soliton solutions}

We have seen in sections \ref{sectwodeg} and \ref{secthreedeg} that the
individual one-soliton constituents within the degenerate two and
three-soliton solutions (\ref{utwo}) and (\ref{uthree}) are shifted relative
to each other by the time-dependent displacements (\ref{delt}) and (\ref%
{delt2}), respectively. As these shifts are logarithmic in time the change
is very slow and when confined to some finite regions they may be viewed as
a compound $N$-soliton as advocated in \cite{FranciscoAndreas}. This
qualitative behaviour remains the same for degenerate $N$-soliton solutions
for any $N$, albeit the fomulae for the time-dependent displacements (\ref%
{delt}) and (\ref{delt2}) need to be generalized.

Using the notation%
\begin{equation}
\lim_{\alpha _{2},\ldots ,\alpha _{N}\rightarrow \alpha _{1}=\alpha
}u_{i\theta _{1}=i\theta ,\ldots ,i\theta _{N};\alpha _{1},\ldots ,\alpha
_{N}}(x,t)=p_{i\theta ,\ldots ,i\theta _{N};N\alpha }(x,t)+iq_{i\theta
,\ldots ,i\theta _{N};N\alpha }(x,t)
\end{equation}%
We compute the asymptotic limits for $N$ even and $N$ odd separately. For
the even case we compute 
\begin{eqnarray}
\lim_{t\rightarrow \sigma \infty }p_{i\theta ,\ldots ,i\theta _{2n};2n\alpha
}\left[ t\alpha ^{2}+\sigma \Delta _{n,\ell ,1}(t),t\right] &=&\hat{P}%
_{\alpha }\left( \theta +\frac{1-(-1)^{n+\ell +1}}{2}\pi \right) \, \\
\lim_{t\rightarrow \sigma \infty }p_{i\theta ,\ldots ,i\theta _{2n};2n\alpha
}\left[ t\alpha ^{2}-\sigma \Delta _{n,\ell ,1}(t),t\right] &=&\hat{P}%
_{\alpha }\left( \theta +\frac{1-(-1)^{n+\ell }}{2}\pi \right)
\end{eqnarray}%
for $n=1,2,\ldots $, $\ell =1,2,\ldots ,n$ and for the odd case we obtain 
\begin{equation}
\lim_{t\rightarrow \sigma \infty }p_{i\theta ,\ldots ,i\theta
_{2n+1};(2n+1)\alpha }\left[ t\alpha ^{2}\pm \Delta _{n,\ell ,0}(t),t\right]
=\hat{P}_{\alpha }\left( \theta +\frac{1-(-1)^{n+\ell }}{2}\pi \right)
\end{equation}%
for $n=0,1,2,\ldots $, $\ell =0,1,2,\ldots ,n$. The time-dependent
displacement takes on the general form%
\begin{equation}
\Delta _{n,\ell ,\kappa }(t)=\frac{1}{\alpha }\ln \left[ \frac{(n-\ell )!}{%
(n+\ell -\kappa )!}(4\left\vert t\right\vert \alpha ^{3})^{2\ell -\kappa }%
\right] .  \label{tdd}
\end{equation}%
As in the previous cases we could also track the minima in the real part
when they are present or the minima and maxima in the imaginary part, which
leads to the same expressions for the time-dependent displacement (\ref{tdd}%
).

\section{Reality conditions for conserved charges}

\label{realitycond}

We will now argue that $\mathcal{PT}$-symmetry \textbf{together} with
integrability will guarantee that \textbf{all} conserved charges in the
model will be real. In order to see the structure of all of the charges we
briefly recall how they can be constructed from the so-called Gardner
transformation \cite{miura1968korteweg,Miura,kupershmidt1981nature}. The
central idea is to expand the KdV-field $u(x,t)$ in terms of a new field $%
w(x,t)$%
\begin{equation}
u(x,t)=w(x,t)+\varepsilon w_{x}(x,t)-\varepsilon ^{2}w^{2}(x,t),  \label{uw}
\end{equation}%
for some deformation parameter $\varepsilon \in \mathbb{R}$. The
substitution of $u(x,t)$ into the KdV equation (\ref{KdV}) yields%
\begin{equation}
\left( 1+\varepsilon \partial _{x}-2\varepsilon ^{2}w\right) \left[
w_{t}+\left( w_{xx}+3w^{2}-2\varepsilon ^{2}w^{3}\right) _{x}\right] =0.
\end{equation}%
Since the last bracket is in form of a conservation law and needs to vanish
by itself, one concludes that $\dint\nolimits_{-\infty }^{\infty }w(x,t)dx=~$%
const. Expanding the new field as%
\begin{equation}
w(x,t)=\dsum\limits_{n=0}^{\infty }\varepsilon ^{n}w_{n}(x,t)
\end{equation}%
then implies that also the quantities $I_{n}:=\dint\nolimits_{-\infty
}^{\infty }w_{2n-2}(x,t)dx$ are conserved. We may then use the relation (\ref%
{uw}) to construct the charge densities in a recursive manner%
\begin{equation}
w_{n}=u\delta _{n,0}-\left( w_{n-1}\right)
_{x}+\dsum\limits_{k=0}^{n-2}w_{k}w_{n-k-2}.  \label{rec}
\end{equation}%
Solving (\ref{rec}) recursively, by taking $w_{n}=0$ for $n<0$, we obtain
easily the well known expressions for the first charge densities%
\begin{eqnarray}
w_{0} &=&u, \\
w_{1} &=&-\left( w_{0}\right) _{x}=-u_{x}, \\
w_{2} &=&-\left( w_{1}\right) _{x}+w_{0}^{2}=u_{xx}+u^{2}, \\
w_{3} &=&-\left( w_{2}\right) _{x}+2w_{0}w_{1}=-u_{xxx}-2(u^{2})_{x}, \\
w_{4} &=&-\left( w_{3}\right)
_{x}+2w_{0}w_{2}+w_{1}^{2}=u_{xxxx}+6(uu_{x})_{x}+2u^{3}-u_{x}^{2}.
\end{eqnarray}%
The expressions simplify substantially when we drop surface terms and we
recover the first three charges reported in (\ref{mpe}). For the energy to
be part of this general series is the reason why we rescaled it as compared
to \cite{CenFring,FranciscoAndreas}.

For the charges constructed from the one-soliton solution (\ref{newS}) we
obtain real expressions%
\begin{equation}
I_{n}=\dint\nolimits_{-\infty }^{\infty }w_{2n-2}(x,t)dx=\frac{2}{2n-1}%
\alpha ^{2n-1}\qquad \text{and\qquad }I_{n/2}=0.  \label{III}
\end{equation}%
The reality of all charges build on one-soliton solutions is guaranteed by $%
\mathcal{PT}$-symmetry alone: When realizing the $\mathcal{PT}$-symmetry as $%
\mathcal{PT}$: $u\rightarrow u$, $x\rightarrow -x$, $t\rightarrow -t$, $%
i\rightarrow -i$ it is easily seen from (\ref{rec}) that the charge
densities transform as $w_{n}\rightarrow (-1)^{n}w_{n}$. This mean when $%
u(x,t)$ is $\mathcal{PT}$-symmetric so are the even graded charge densities $%
w_{2n}(x,t)$. Changing the argument of the functional dependence to the
traveling wave coordinate $\zeta _{\alpha }=x-\alpha ^{2}t$ this means we
can separate $w_{2n}(\zeta _{\alpha })$ into a $\mathcal{PT}$-even and $%
\mathcal{PT}$-odd part $w_{2n}^{e}(\zeta _{\alpha })\in \mathbb{R}$ and $%
w_{2n}^{o}(\zeta _{\alpha })\in \mathbb{R}$, respectively, as $w_{2n}(\zeta
_{\alpha })=w_{2n}^{e}(\zeta _{\alpha })+iw_{2n}^{o}(\zeta _{\alpha })$,
which allows us to conclude%
\begin{equation}
I_{n}(\alpha )=\dint\nolimits_{-\infty }^{\infty
}w_{2n-2}(x,t)dx=\dint\nolimits_{-\infty }^{\infty }\left[
w_{2n-2}^{e}(\zeta _{\alpha })+iw_{2n-2}^{o}(\zeta _{\alpha })\right] d\zeta
_{\alpha }=\dint\nolimits_{-\infty }^{\infty }w_{2n-2}^{e}(\zeta _{\alpha
})d\zeta _{\alpha }\in \mathbb{R}.
\end{equation}%
It is easily seen that the previous argument applies directly to the charges
build from the solution $u_{i\theta ;\alpha }(x,t)$ in (\ref{newS}), i.e.
the real part and imaginary part are even and odd in $\zeta _{\alpha }$,
respectively. When the parameter $\mu $ has a nonvanishing real part the $%
\mathcal{PT}$-symmetry is broken, but it can be restored by absorbing the
real part by a shift either in $t$ or $x$ as argued in \cite{CenFring}.

In order to ensure the same for the multi-soliton solutions we use the fact
that the multi-soliton solutions separate asymptotically into single
solitons with distinct support. As the charges are conserved in time we may
compute $I_{n}$ at any time. In the asymptotic regime any charge build from
an $N$-soliton $u_{i\theta _{1},\ldots ,i\theta _{N};\alpha _{1},\ldots
,\alpha _{N}}^{(N)}$ decomposes into the sum of charges build on the
one-soliton solutions. 
\begin{eqnarray}
I_{n}(\alpha _{1},\ldots ,\alpha _{N}) &=&\dint\nolimits_{-\infty }^{\infty
}\left( w_{i\theta _{1},\ldots ,i\theta _{N};\alpha _{1},\ldots ,\alpha
_{N}}^{(N)}\right) _{2n-2}(x,t)dx,  \label{I1} \\
&=&\dint\nolimits_{-\infty }^{\infty }\dsum\nolimits_{k=1}^{N}\left[ \left(
w_{i\theta _{k};\alpha _{k}}^{(1)}\right) _{2n-2}(\zeta _{\alpha _{k}})%
\right] d\zeta _{\alpha _{k}},  \label{I2} \\
&=&\dsum\nolimits_{k=1}^{N}I_{n}(\alpha _{k}),  \label{I3} \\
&=&\frac{2}{2n-1}\dsum\nolimits_{k=1}^{N}\alpha _{k}^{2n-1}.  \label{I4}
\end{eqnarray}%
We used here the decomposition of the N-soliton into a sum of one-solitons
in the asymptotic regime $u_{i\theta _{1},\ldots ,i\theta _{N};\alpha
_{1},\ldots ,\alpha _{N}}^{(N)}=\dsum\nolimits_{k=1}^{N}\left( u_{i\theta
_{k};\alpha _{k}}^{(1)}\right) $, which we have seen in detail above. Since
each of the one-solitons is well localized we always have $u_{i\theta
_{k};\alpha _{k}}^{(1)}\cdot u_{i\theta _{l};\alpha _{l}}^{(1)}=0$ when $%
k\neq l$, which implies that 
\begin{equation}
\left[ u_{i\theta _{1},\ldots ,i\theta _{N};\alpha _{1},\ldots ,\alpha
_{N}}^{(N)}\right] ^{m}=\left[ \dsum\nolimits_{k=1}^{N}\left( u_{i\theta
_{k};\alpha _{k}}^{(1)}\right) \right] ^{m}=\dsum\nolimits_{k=1}^{N}\left(
u_{i\theta _{k};\alpha _{k}}^{(1)}\right) ^{m}.
\end{equation}%
As all the derivatives are finite and the support is the same as for the $u$%
s, this also implies 
\begin{equation}
\left[ \left( u_{i\theta _{1},\ldots ,i\theta _{N};\alpha _{1},\ldots
,\alpha _{N}}^{(N)}\right) _{nx}\right] ^{m}=\left[ \dsum\nolimits_{k=1}^{N}%
\left( u_{i\theta _{k};\alpha _{k}}^{(1)}\right) _{nx}\right]
^{m}=\dsum\nolimits_{k=1}^{N}\left( u_{i\theta _{k};\alpha
_{k}}^{(1)}\right) _{nx}^{m},
\end{equation}%
and similarly for mixed terms involving different types of derivatives. As
all charge densities are made up from $u$ and its derivatives we obtain%
\begin{equation}
\left( w_{i\theta _{1},\ldots ,i\theta _{N};\alpha _{1},\ldots ,\alpha
_{N}}^{(N)}\right) _{2n-2}=\dsum\nolimits_{k=1}^{N}\left( w_{i\theta
_{k};\alpha _{k}}^{(1)}\right) _{2n-2}
\end{equation}%
in the asymptotic regime, which is used in the step from (\ref{I1}) to (\ref%
{I2}). In the remaining two steps (\ref{I3}) and (\ref{I4}) we use (\ref{III}%
).

Thus $\mathcal{PT}$-symmetry \textbf{and} integrability guarantee the
reality of \textbf{all} charges.

\section{Conclusions}

We have explicitly computed lateral displacements and time-delays for
complex two and three-soliton solutions of the KdV equation. Our solutions
satisfy the consistency equations (\ref{SM}) and (\ref{SP}) resulting from
the preservation of the centre of mass coordinate. The expressions for the
time-delay of the three-soliton scattering, being the sum of the delays of
two two-soliton time-delays, confirm on classical level the standard
factorization property of the scattering matrix for integrable systems that
allows to treat any multiple scattering process as a succession of two
particle scatterings. The imaginary part in our solutions may be thought of
as a massless soliton partaking in the scattering process.

We used our expressions for three different purposes: Firstly we made the
relation between solutions obtained from Hirota's direct method on one hand
and those constructed from a superposition principle based on B\"{a}cklund
transformations precise. They differ by non-identical lateral displacements
in each of their one-soliton constituents and additional shifts by $\pi $ in
the shift parameters. Overall they lead to the same values of the
time-delays as they are preserved quantities. Secondly we elaborated on the
internal structure of compound soliton solutions within degenerate
multi-soliton solutions. We found that the degenerate one-soliton
contributions are displaced relative to each other by a time-dependent
shift. When scattered with any nondegenerate one-soliton constituent they
are all displaced by the same amount. With (\ref{tdd}) we presented a
generic formula for the relative time-dependent displacement valid for any
degenerate $N$-soliton solution. Thirdly we clarified the role $\mathcal{PT}$%
-symmetry plays in guaranteeing the reality of conserved charges, the energy
being one of them. It turned out that $\mathcal{PT}$-symmetry is solely
responsible for the reality of any charge based on one-soliton solutions.
For charges constructed from multi-soliton solutions we need to invoke
integrability having the effect of separating asymptotically the
multi-solitons into single solitons to ensure the reality these charges.

As our approach is entirely model independent, it would naturally be
interesting to apply it to other complex integrable systems. Furthermore, it
would be very interesting to employ the expressions obtained for a
semi-classical quantization. In particular the role played here by the
massless soliton might shed some new light on some old results \cite%
{zamomassless}.

\medskip \noindent \textbf{Acknowledgments:} FC would like to thank the
Alexander von Humboldt Foundation (grant number CHL 1153844 STP) for
financial support and City University London for kind hospitality.

\appendix

\section{Sample time-delay computations}

\label{AppendixA} Most of the computations are rather cumbersome, so that it
suffices to present a few samples. Let us for instance derive the values for
the shifts in (\ref{2real}). For simplicity we take the real part of $%
~u_{i\theta ,i\phi ;\alpha ,\beta }(x,t)$ in (\ref{2S}) at specific values
of $\theta $ and $\phi $. Taking $\theta =\phi =\pi /2$ it acquires the
relatively simple form 
\begin{equation}
\!\frac{4\left[ \alpha e^{\alpha ^{3}t+2\beta ^{3}t+\alpha x}+\beta
e^{2\alpha ^{3}t+\beta ^{3}t+\beta x}+\alpha \chi (\alpha ,\beta )e^{\alpha
^{3}t+\alpha x+2\beta x}+\beta \chi (\alpha ,\beta )e^{\beta ^{3}t+2\alpha
x+\beta x}\right] ^{2}}{\left[ e^{2t\left( \alpha ^{3}+\beta ^{3}\right) }+%
\frac{8\alpha \beta }{(\alpha +\beta )^{2}}e^{t\left( \alpha ^{3}+\beta
^{3}\right) +x(\alpha +\beta )}+e^{2\alpha ^{3}t+2\beta x}+e^{2\beta
^{3}t+2\alpha x}+\chi ^{2}(\alpha ,\beta )e^{2x(\alpha +\beta )}\right] ^{2}}%
.  \label{app0}
\end{equation}%
Tracking the soliton with speed $\alpha ^{2}$ in the multi-soliton solution
and trying to match it with the one-soliton, we may try to solve the
equivalence relation 
\begin{equation}
u_{i\pi ,i\pi ;\alpha ,\beta }(t\alpha ^{2}+\Delta ,t)\sim u_{i\pi ;\alpha
}(t\alpha ^{2},t),  \label{app1}
\end{equation}%
in the asymptotic regimes $t\rightarrow \pm \infty $ for some as yet unknown
constant $\Delta $. The right hand side of (\ref{app1}) is easily computed
from (\ref{newS}) to $\alpha ^{2}$. Replacing $x$ by $t\alpha ^{2}+\Delta $
in (\ref{app0}) and identifying $e^{4t(\alpha ^{3}+\alpha ^{2}\beta )}$ as
the dominant term in the numerator and denominator in the large $t$ regime
we read off the corresponding coefficients. Thus for large $t$ the
equivalence relation (\ref{app1}) becomes

\begin{equation}
\frac{4\alpha ^{2}\left( \alpha ^{2}-\beta ^{2}\right) ^{4}e^{2\alpha \Delta
}}{2\left( \alpha ^{2}-\beta ^{2}\right) ^{4}e^{2\alpha \Delta }+(\alpha
-\beta )^{8}e^{4\alpha \Delta }+(\alpha +\beta )^{8}}=\alpha ^{2}.
\end{equation}%
Solving this equation for $\Delta $ leads to the time-displacement $\Delta
=2/\alpha \ln [(\alpha +\beta )/(\alpha -\beta )]$ for $t\rightarrow \infty $%
, which corresponds to the shift reported in (\ref{2real}) for generic
values of $\theta $ and $\phi $.

Next we present a sample computation for a degenerate multi-soliton solution
for which the lateral displacement becomes time dependent rather than just
being constant. We derive the asymptotic relation (\ref{s1}). Taking $x$ to
be $t\alpha ^{2}+\Delta $ in the degenerate two-soliton solution (\ref{utwo}%
) for some constant $\Delta $, we observe that the limits $t\rightarrow \pm
\infty $ always yield zero and we will not be able to obtain a finite value
such as the maximum of the one-soliton $\hat{P}_{\alpha }(\theta )$. Hence
we are forced to include a time-dependence into $\Delta $. This mildest
dependence we may introduce is a logarithmic one. Taking therefore as an
Ansatz $\Delta (t)=-1/\alpha \ln (\kappa \left\vert t\right\vert )$ for some
unknown constant $\kappa $ we compute%
\begin{equation}
\frac{4\alpha ^{2}e^{i\theta }\kappa \left\vert t\right\vert \left[ \kappa
^{2}t^{2}\left( 2\alpha ^{3}t-\nu -2+\ln (\kappa \left\vert t\right\vert
)\right) -e^{2i\theta }\left( 2\alpha ^{3}t+2-\nu +\ln (\kappa \left\vert
t\right\vert )\right) -4e^{i\theta }\kappa \left\vert t\right\vert \right] }{%
\left[ e^{2i\theta }-\kappa ^{2}t^{2}-2e^{i\theta }\kappa \left\vert
t\right\vert \left( 2\alpha ^{3}t-\nu +\ln (\kappa \left\vert t\right\vert
)\right) \right] ^{2}}.  \label{app4}
\end{equation}%
For large $\left\vert t\right\vert $ we can now find matching powers in $t$
in the numerator and denominator. We replace now $\left\vert t\right\vert $
by $\sigma t$ and take $\sigma $ to be $\pm 1$ depending on whether $t$ is
negative or positive. The leading order terms are proportional to $t^{4}$.
Neglecting all other terms,\ the expression in (\ref{app4}) reduces to%
\begin{equation}
\frac{8\alpha ^{5}e^{i\theta }\kappa \sigma }{\left( \kappa +4\sigma \alpha
^{3}e^{i\theta }\right) ^{2}}.
\end{equation}%
For $\kappa =$\ $4\alpha ^{3}$ this equals $\hat{P}_{\alpha }(\theta )$ and $%
\hat{P}_{\alpha }(\theta +\pi )$ for $\sigma =1$ and $\sigma =-1$,
respectively. Thus we have derived (\ref{s1}) and (\ref{s2}) with the
lateral displacement takeing on the form (\ref{delt}).

\newif\ifabfull\abfulltrue

%%\bibliographystyle{phreport}
%%\bibliography{acompat,Ref}

\end{document}